\documentclass[preprint,12pt]{elsarticle}

\usepackage{hyperref}
\usepackage{graphicx}
\usepackage{subcaption}
\usepackage{amssymb}
\usepackage{amsmath}
\usepackage{multirow}
\usepackage[utf8]{inputenc}

\usepackage{xcolor}
\usepackage{xargs}
\usepackage[colorinlistoftodos,textsize=footnotesize]{todonotes}
\usepackage{lineno}
\usepackage{cleveref}
\usepackage{algorithm}
\usepackage[noend]{algpseudocode}
\usepackage{listings}
\usepackage[section]{placeins}
\usepackage{microtype}

\newcommandx{\unsure}[2][1=]{\todo[linecolor=red,backgroundcolor=red!25,bordercolor=red,#1]{#2}}
\newcommandx{\change}[2][1=]{\todo[linecolor=blue,backgroundcolor=blue!25,bordercolor=blue,#1]{#2}}
\newcommandx{\info}[2][1=]{\todo[linecolor=OliveGreen,backgroundcolor=OliveGreen!25,bordercolor=OliveGreen,#1]{#2}}

\newcommand{\ten}[1]{\ensuremath{\mathbf{#1}}}

\journal{}

\begin{document}

\begin{frontmatter}

  \title{A Second-Order, Variable-Resolution, Weakly-Compressible Smoothed Particle Hydrodynamics Scheme}
  \author[IITB]{Abhinav Muta\fnref{fn1}}
  \author[IITB,IIT]{Pawan Negi\corref{cor1}\fnref{fn1}}
  \ead{pawan.n@aero.iitb.ac.in}
  \author[IITB]{Prabhu Ramachandran}
  \ead{prabhu@aero.iitb.ac.in}
\address[IITB]{Department of Aerospace Engineering, Indian Institute of
  Technology Bombay, Powai, Mumbai 400076}
\address[IIT]{Department of Applied Mathematics, Illinois Institute of Technology,
  Chicago, 60616}

\cortext[cor1]{Corresponding author}
\fntext[fn1]{Joint first author}

\begin{abstract}
The smoothed particle hydrodynamics (SPH) method has been widely used to
simulate incompressible and slightly compressible fluid flows. Adaptive
refinement strategies to dynamically increase the resolution of the particles
to capture sharp gradients in the flow have also been developed. However, most
of the SPH schemes in the literature are not second-order convergent (SOC).
Both second-order convergence and adaptive resolution are considered grand
challenge problems in the SPH community. In this paper, we propose, for the
first time, a second-order convergent (SOC) adaptive refinement strategy along
with a SOC weakly-compressible SPH scheme. We employ the method of
manufactured solutions to systematically develop the scheme and validate the
solver. We demonstrate the order of convergence of the entire scheme using the
Taylor-Green vortex problem and then go on to demonstrate the applicability of
the method to simulate flow past a circular cylinder.
  \end{abstract}


\end{frontmatter}

% \linenumbers

\section{Introduction}
\label{sec:intro}

\footnote{© 2024. This manuscript version is made available under the
CC-BY-NC-ND 4.0 license
\url{https://creativecommons.org/licenses/by-nc-nd/4.0/}}
The smoothed particle hydrodynamics (SPH) method is a Lagrangian, meshless
method widely used to simulate continuum mechanics problems including fluid
dynamics~\cite{monaghan-review:2005,edac-sph:cf:2019,sunConsistent2019} and
elastic dynamics~\cite{zhang_hu_adams17,AdepuCTVF2023}. The meshless nature of the method
inherently allows adaptivity of the particles according to the dynamics of the
specific problem. However, there are still several
challenges~\cite{vacondio_grand_2020} to overcome for SPH to emerge as a
reliable equivalent to traditional mesh-based methods applied in industry.

As discussed by \citet{vacondio_grand_2020}, a second-order convergent
Lagrangian SPH scheme is a grand-challenge problem. The root cause of
non-convergence is the selection of a non-convergent (but conservative)
discretization method~\cite{NegiTechniques2022}. In SPH, a discretization that
is both convergent and conservative is not
available~\cite{priceSmoothedParticleHydrodynamics2012,NegiTechniques2022}.
Therefore, one has to choose between a conservative or a convergent
discretization. According to \citet{priceSmoothedParticleHydrodynamics2012},
conservation is important, furthermore it provides inherent particle
regularization. However, \citet{NegiTechniques2022} show that in the case of
incompressible and weakly-compressible fluids, the linear momentum remains
bounded (though not conserved) when one uses a convergent scheme and the
accuracy achieved is very high. Along similar lines, many authors
\cite{hashemiModifiedSPHMethod2012,ravanbakhshImplementationImprovedSpatial2023,NegiTechniques2022,negiHowTrainYour2022,negiHowTrainYour2021}
have proposed convergent techniques that are not conservative. In order to
construct a second-order convergent scheme, a kernel gradient
correction~\cite{bonet_lok:cmame:1999,liuRestoringParticleConsistency2006} of
some form is necessary and this can be computationally expensive. 

Traditional SPH schemes for incompressible or weakly-compressible fluids have
tended to employ a fixed resolution of particles wherein all particles are of
the same spatial resolution. As can be expected, this is highly inefficient
especially in the case of fluid dynamics where high-resolution is only
necessary in certain regions. As a result, many adaptive resolution schemes
have been proposed over the years where the particle resolution is adaptively
refined.

\citet{feldman_dynamic_2007} proposed a dynamic refinement strategy ensuring
mass conservation. They ensured that the global kinetic energy, linear
momentum, and angular momentum are conserved by the method.
\citet{vacondio_accurate_2012,vacondio2013variable} developed this further and
proposed a splitting as well as a merging technique ensuring corrections due
to variable particle resolution. \citet{barcarolo_adaptive_2014} proposed to
retain the parent particle after splitting to use it as a support while
merging.  \citet{chiron_apr_2018} improved the adaptive particle refinement
proposed by \citet{vacondio_splitting_2013,vacondio_variable_2016}. Their
proposed method resolved issues due to particles of different sizes
interacting and retained the parent and child particles in the region between
two different resolutions.
\citet{sun_plus-sph_2017,sun_multi-resolution_2018,sun2021a} also used this
approach with a buffer region between two distinct particle resolutions. Two
different resolution do not interact while computing the accelerations.
Recently, \citet{ricciMultiscaleSmoothedParticle2024} used a similar approach
to decompose the domain into different layers that interact through buffer
layers. \citet{yang_smoothed_2017,yang_adaptive_2019} proposed a dynamic
strategy to set the smoothing length that allowed a continuous variation of
the smoothing length from minimum to maximum resolution in the domain.
Recently, \citet{mutaAdaptive2022}, followed by an improvement in
\citet{haftuParallelAdaptiveWeaklycompressible2022}, proposed a parallel
implementation that combines the approaches of \citet{vacondio_variable_2016}
along with that of \citet{yang_adaptive_2019}.  The method is accurate, has
optimal smoothing lengths, requires much fewer particles than other schemes,
does not require a manual specification of spatial resolution, and also fully
supports object motion and solution adaptivity.  We note that none of the
methods discussed require a support mesh to determine the particle
resolutions. While these methods allows accurate solutions to the shallow
water equations~\cite{vacondio_accurate_2012}, soil
simulation~\cite{reyeslopez2013}, fluid mechanics
problems~\cite{haftuParallelAdaptiveWeaklycompressible2022}, fluid-structure
interaction~\cite{hu2019} and multi-phase
simulations~\cite{yang_adaptive_2019}. However, none of the above methods are
second-order accurate. As mentioned earlier, this is a grand-challenge
problem.

In this paper, we propose a second-order convergent (SOC) adaptive refinement
strategy and a SOC scheme for variable smoothing length
particles. We develop the method in two parts. We first develop a SOC scheme
for variable smoothing length. We use the first-order consistency correction
for the kernel gradients in the approximation of all the operators. This
ensures correct estimation of the property irrespective of the smoothing
length variation. We use particle shifting technology
(PST)~\cite{diff_smoothing_sph:lind:jcp:2009} to iteratively shift particle
after every few iterations. We use the method of manufactured solutions (MMS)
as proposed in \cite{negiHowTrainYour2021} to ensure that the new scheme is
indeed SOC in the presence of variable smoothing length.

In the second part, we use the adaptive refinement strategy proposed by
\citet{haftuParallelAdaptiveWeaklycompressible2022}. The adaptive strategy
consist of three parts where the field properties are updated, namely,
particle splitting, particle merging, and shifting. We propose a second-order
convergent approximation for each of these updates and use the original
algorithm as is. Once again, we use the MMS to show that the adaptive
refinement strategy is indeed SOC.

In the next section, we briefly discuss the SPH method and the types of
approximation that works in the presence of variable smoothing length, $h$. In
\cref{sec:soc_edac}, we discuss the second-order scheme followed by the
second-order convergent adaptive refinement strategy. We discuss the original
algorithm that we do not change in the \ref{sec:adaptive-soc_org}. In
\cref{sec:results}, we demonstrate the accuracy of the proposed method by
simulating Taylor-Green vortex and the flow past a circular cylinder. In the
interest of reproducibility we automate~\cite{pr:automan:2018} the results of
this work. The entire source-code of the work conducted in this paper can be
found at \url{https://gitlab.com/pypr/adaptive_mms}.

\section{The SPH method}
\label{sec:sph}

In the SPH method~\cite{monaghan-gingold-stars-mnras-77}, the scalar field $f$
defined in the domain $\Omega$ with boundary $\partial \Omega$ is
reconstructed (away from the domain boundary), with second-order accuracy,
using a smoothing kernel $W$ as
\begin{equation}
  f(\ten{x}) = \int_\Omega f(\tilde{\ten{x}}) W(\ten{x} - \tilde{\ten{x}},
  h) d \tilde{\ten{x}} + O(h^2),
\end{equation}
where $\ten{x} \in \Omega$ and $h$ is the smoothing length of the
kernel. Similarly, the gradient of
the scalar field
\begin{equation}
  \nabla f(\ten{x}) = \int_\Omega f(\tilde{\ten{x}}) \nabla W(\ten{x} -
  \tilde{\ten{x}}, h) d \tilde{\ten{x}} + O(h^2).
\end{equation}
The domain can be discretized using particles having uniform mass and constant
smoothing length. Using discrete particles, the SPH discretization of the
function gradient
\begin{equation}
  \nabla f(\ten{x}_i) = \sum_{j=1}^N f(\ten{x}_j) \nabla W_{ij} \omega_j + E_{\text{quad}} + O(h^2),
\end{equation}
where $E_{\text{quad}}$ is the quadrature error
\cite{quinlan_truncation_2006}, $\omega_j$ is the volume of the particle,
$\nabla W_{ij}= \nabla_i W(\ten{x}_i - \ten{x}_j, h_{ij})$, and the sum is
taken over $N$ neighbors of the particle $i$. We note that it is possible to
employ a distribution of particles with variable mass $m$ and variable
smoothing length $h$. \citet{vacondio2013variable} proposed a formulation to
evaluate the gradient of a scalar field to account for the variation of $h$ as
\begin{equation}
  \left< \nabla f(\ten{x}_i) \right> = \sum_j m_i m_j \left( \frac{f_j}{\beta_j \rho_j^2}
  \nabla W_{j} -  \frac{f_i}{\beta_i \rho_i^2}
  \nabla W_i \right),
  \label{eq:conv_adapt_grad}
\end{equation}
where, 
\begin{equation}
  \beta_i = - \frac{1}{\psi_i d} \sum_j m_j \ten{x}_{ij} \nabla W_{ij},
\end{equation}
is the correction term due to the variation in $h$, $\psi_i$ is the
particle density, $d$ is the dimension of the problem, $\ten{x}_{ij} =
\ten{x}_i - \ten{x}_j$, and $\nabla W_i =\nabla W(\ten{x}_i - \ten{x}_j, h_{i})$.
The formulation in \cref{eq:conv_adapt_grad} conserves linear momentum
pair-wise. In the context of the SPH method, many researchers
\cite{priceSmoothedParticleHydrodynamics2012,NegiTechniques2022} have shown
that a formulation that is conservative and second-order convergent is not
available. Recently, \citet{sun_accurate_2021, luthi_adaptive_2023,
pearlFSISPHSPHFormulation2022} employed kernel gradient correction to
correctly evaluate gradients in the presence of variable resolution. Using a
similar technique, we propose a second-order convergent gradient estimation, in
the presence of variable resolution,
\begin{equation}
  \left< \nabla f(\ten{x}_i) \right> = \sum_j f_j \nabla \tilde{W}_j \omega_j,
  \label{eq:adapt_grad_scalar}
\end{equation}
where $\nabla \tilde{W}_j = L_i \nabla W(\ten{x}_i - \ten{x}_j, h_j)$ is the
corrected kernel gradient, where $L_i$ is the first order correction matrix
proposed by \citet{liuRestoringParticleConsistency2006} and discussed in
greater detail in \cref{sec:edac}. In the next section, we extend this
formulation and propose a second-order convergent scheme for an adaptively
refined domain.

\section{A second-order convergent scheme for variable $h$}
\label{sec:soc_edac}

The time-accurate simulation of an incompressible fluid can be performed using
the weakly-compressible Navier-Stokes (N-S)~\cite{sph:fsf:monaghan-jcp94}
equations
\begin{equation}
  \begin{split}
    \frac{d \rho}{ d t} &= -\rho \nabla \cdot \ten{u}, \\
    \frac{d \ten{u}}{d t} &= \ -\frac{\nabla p}{\rho} + \nu \nabla^2 \ten{u},
  \end{split}
  \label{eq:wcsph}
\end{equation}
where $\rho$ and $\nu$ are the fluid density and kinematic viscosity,
respectively.  The above equation is closed using an artificial equation of
state
\begin{equation}
  p = c_o^2 ( \rho - \rho_o),
  \label{eq:eos}
\end{equation}
where $c_o$ is the artificial speed of sound and $\rho_o$ is the reference
density. Using the equation of state and the entropically damped artificial
compressibility (EADC)
approach~\cite{Clausen2013,Clausen2013a,edac-sph:cf:2019}, the continuity
equation can be written as a pressure evolution equation
\begin{equation}
  \frac{d p}{ d t} = -\rho c_o^2 \nabla \cdot \ten{u} + D \nabla^2 p,
  \label{eq:pres_evol}
\end{equation}
where $D=\alpha h c_o / 8$ is the density damping coefficient, where is
recommended to be $\alpha=0.5$ \cite{edac-sph:cf:2019}.

In this work, we extend the existing scheme proposed by
\citet{mutaAdaptive2022} using the pressure evolution
\cref{eq:pres_evol} along with the momentum equation to simulate fluid flows.
However, we may also choose to use a density evolution equation and this has
no impact on the convergence of the method as demonstrated
in~\cite{NegiTechniques2022}. We divide the construction of the proposed
scheme in two stages. We first propose a SOC scheme in the presence of
variable resolution. We use a particle shifting technique
\cite{diff_smoothing_sph:lind:jcp:2009} to ensure uniform particle
distribution with an accurate Taylor series correction. In the subsequent
stage, we add a SOC splitting and merging to the scheme. At each stage, we use
the method of manufactured solutions proposed by \citet{negiHowTrainYour2021}
to verify the convergence. We employ the techniques proposed in
\cite{NegiTechniques2022} to construct the discrete operators. The boundary
(solid, inlet, outlet) particles have the same resolution as the nearby fluid
particle. Therefore, we can directly apply the convergent boundary condition
implementation identified in \cite{negiHowTrainYour2022}. In the next
section, we describe the construction of a second-order convergent
EDAC scheme.

\subsection{Second-order EDAC scheme}
\label{sec:edac}

\citet{NegiTechniques2022} proposed a SOC scheme with pressure evolution for a
uniform resolution.  However, in the case of variable resolution, a
straight-forward implementation does not work despite the use of a kernel
gradient correction. In the presence of variable resolution, we set the
particle volume $\omega$ and the corresponding smoothing length $h$ such that
the summation density $\psi$ produces the desired reference particle density,
which in this work is set to $1$. We set the $h$ of a particle using
\begin{equation}
  h_i = C \left( \frac{m_{avg}}{\psi_o}\right)^{1/d},
\end{equation}
where $m_{avg} = \sum_j m_j / N$, $C=1.2$, $d$ is the dimensionality of the
problem, $\psi_o=1$ is the desired particle density (or summation density).
In the case of variable smoothing length, there are two different ways to
compute volume namely, scatter and gather
formulation~\cite{hernquist1989}. \citet{mutaAdaptive2022} proposed to use
gather formulation to compute volume
\begin{equation}
  \omega_i = \frac{1}{\sum_j W(\ten(x_i) - \ten{x}_j, h_i)} = \frac{1}{\sum_j W_i}.
\end{equation}

We use \cref{eq:adapt_grad_scalar} to obtain the
discretization of the continuity equation
\begin{equation}
  \frac{d p_i}{ dt } = -\rho_i c_0^2 \sum_j \ten{u}_j \cdot \nabla \tilde{W}_j
  \omega_j + \frac{\alpha h_i c_o}{8} \sum_j (p_j - p_i) \frac{\ten{x}_{ij} \cdot
  \nabla \tilde{W}_j}{|\ten{x}_{ij}| + 0.01 h_j^2} \omega_j
  \label{eq:cont_disc}
\end{equation}
where $\nabla \tilde{W}_j = L_i \nabla W_j$, and the correction matrix $L_i$
is obtained by solving
\begin{equation}
  \begin{bmatrix}
    \sum_j W_j \omega_j &\sum_j \ten{x}_{ji} W_j \omega_j \\
    \sum_j \ten{x}_{ji} W_j \omega_j &\sum_j (\ten{x}_{ji} \otimes \ten{x}_{ji}) W_j \omega_j \\
  \end{bmatrix}
  \begin{bmatrix}
    \tilde{W}_i \\
    \nabla \tilde{W}_i
  \end{bmatrix}=
  \begin{bmatrix}
    {W}_i \\
    \nabla {W}_i
  \end{bmatrix}.
\end{equation}
We note that the construction of the correction matrix also involves one-sided
kernel function values. The second term in \cref{eq:cont_disc} is a
discretization proposed by \cite{clearyConductionModellingUsing1999} for the
Laplacian term. We do not employ a discretization using a repeated application
of the SOC gradient and divergence operator as it is unstable in the presence
of oscillations \cite{biriukovStableAnisotropicHeat2019}. However, the
presence of the $h_i$ in the damping term makes the error due to the
non-convergent discretization at least first order.

We use a similar discretization for the momentum equation
\begin{equation}
  \frac{d \ten{u}_i}{ d t} = - \frac{1}{\rho_i}\sum_j p_j \nabla \tilde{W}_j
\omega_j + \nu \sum_j \left< \nabla u_i \right> \cdot \nabla \tilde{W}_j
\omega_j,
  \label{eq:disc-momentum}
\end{equation}
where $\left< \nabla u_i \right> = \sum_j \ten{u}_j \otimes \nabla \tilde{W}_j
\omega_j$ is the SOC velocity gradient. The discretized equations
\cref{eq:cont_disc,eq:disc-momentum} are integrated in time using the
second-order Runge-Kutta method. The details of the integrator can be found in
\cite{NegiTechniques2022}.

\begin{figure}[htbp]
  \centering
  \includegraphics[width=\textwidth]{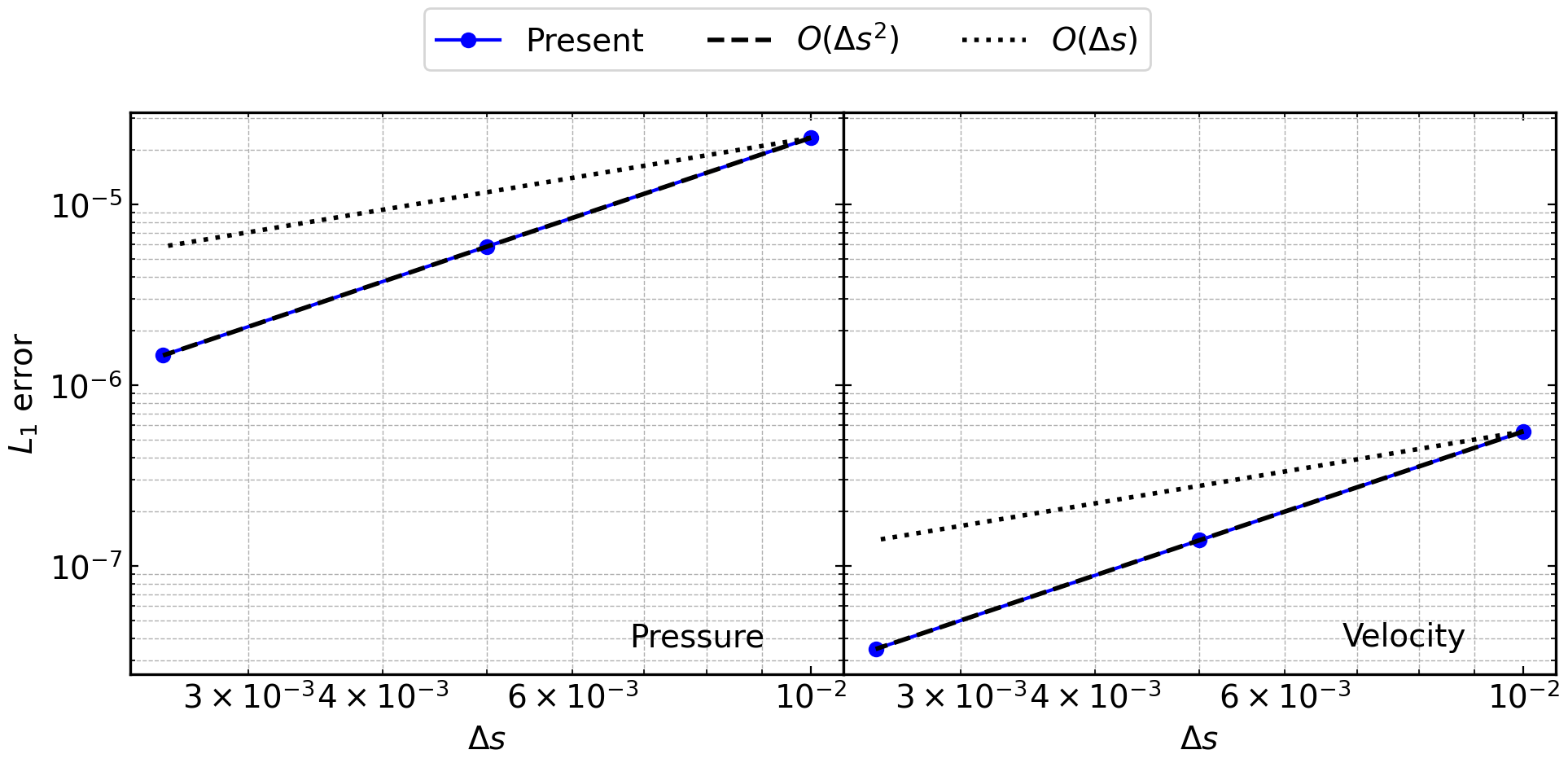}
\caption{$L_1$ error in pressure and velocity after 1 time step using the MS
in \cref{eq:mms_grad_div} with $\alpha=0$ and $\nu=0$.}
  \label{fig:mms_grad_div}
\end{figure}

We use the method of manufactured solutions \cite{negiHowTrainYour2021} to
demonstrate the second-order convergence of the scheme keeping a uniform
resolution. The manufactured solution (MS) used is given by
\begin{equation}
  \begin{split}
    u(x, y) &=\left(y - 1\right) \sin{\left(2 \pi x \right)} \cos{\left(2 \pi y \right)},\\
    v(x, y) &= - \sin{\left(2 \pi y \right)} \cos{\left(2 \pi x \right)},\\
    p(x, y) &= \cos{\left(4 \pi x \right)} + \cos{\left(4 \pi y \right)}.\\
  \end{split}
  \label{eq:mms_grad_div}
\end{equation}
By putting the MS in the governing equation in \cref{eq:wcsph,eq:pres_evol}
with $\nu=0$ and $\alpha=0$, we obtain a source term for momentum and pressure
evolution. The resulting source term is passed to the solver. Since, both the
viscous term and the pressure damping term is zero, the error is only due to
the discretized velocity divergence and pressure gradient operators. In
\cref{fig:mms_grad_div}, we plot the $L_1$ error in pressure and velocity
after 1 time step for the proposed scheme with the change in average particle
spacing $\Delta s$. The rate of convergence is clearly second-order.

\begin{figure}[!htbp]
  \centering
  \includegraphics[width=\textwidth]{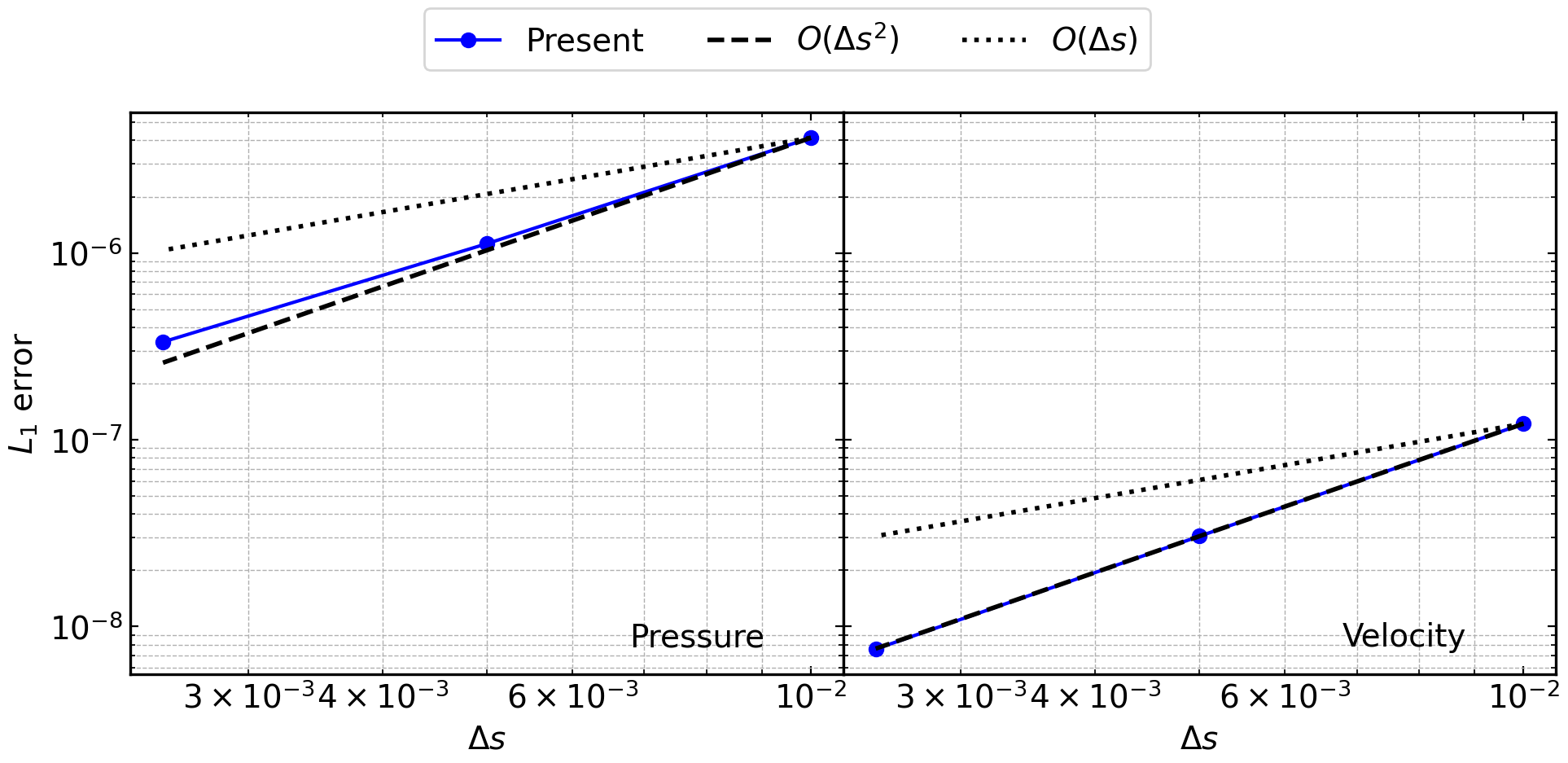}
\caption{$L_1$ error in pressure and velocity after 1 time step using the MS
in \cref{eq:mms_lap} with $\alpha = 0.5$ and $\nu=0.25$.}
  \label{fig:mms_lap}
\end{figure}

We note that the result in \cref{fig:mms_grad_div} demonstrates that the error
may not originate from the velocity divergence and pressure gradient
approximation. In order to test the accuracy of the discretization of the
pressure damping and velocity damping (or the viscous term), we use the MS
given by
\begin{equation}
  \begin{split}
    u(x, y, t) &= \left(y - 1\right) e^{- 10 t} \sin{\left(2 \pi x \right)} \cos{\left(2 \pi
    y \right)},\\
    v(x, y, t) &= - e^{- 10 t} \sin{\left(2 \pi y \right)} \cos{\left(2 \pi x \right)},\\
    p(x, y, t) &= \left(\cos{\left(4 \pi x \right)} + \cos{\left(4 \pi y \right)}\right) e^{-
    10 t}.\\
  \end{split}
  \label{eq:mms_lap}
\end{equation}
We evaluate the source term by setting $\nu=0.25$ and $\alpha=0.5$ in the
governing equation. In \cref{fig:mms_lap}, we plot the $L_1$ error for
pressure and velocity after 1 time step. A slight deviation
from the second-order rate of convergence is visible. It is due to the
non-convergent discretization employed in the pressure gradient term to
mitigate pressure oscillations.

\begin{figure}[!htbp]
  \centering
  \includegraphics[width=\linewidth]{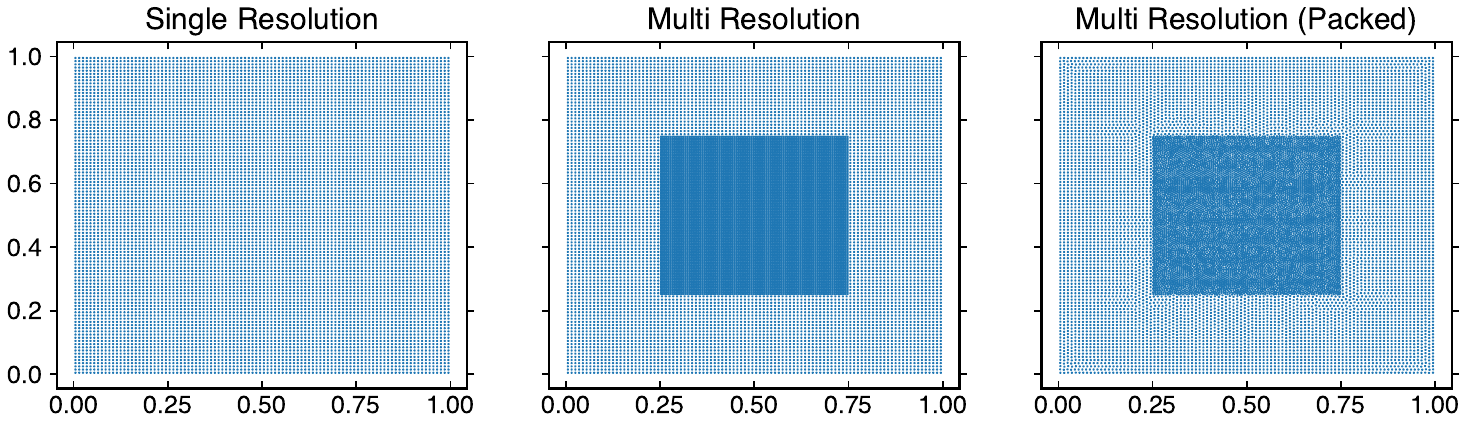}
  \caption{Different particle resolution considered.}
  \label{fig:diff_res}
\end{figure}

In order to test the convergence of the proposed scheme in the presence of
variable resolution, we consider domain with variable resolution as shown in
\cref{fig:diff_res}. We consider a patch with a doubled particle resolution at
the middle of the original uniform particle distribution. We also consider the
patch having non-uniform, packed distribution. In order to perform a convergence
study, we reduce the resolution by half for all particles.

\begin{figure}[!htbp]
  \centering
  \includegraphics[width=\textwidth]{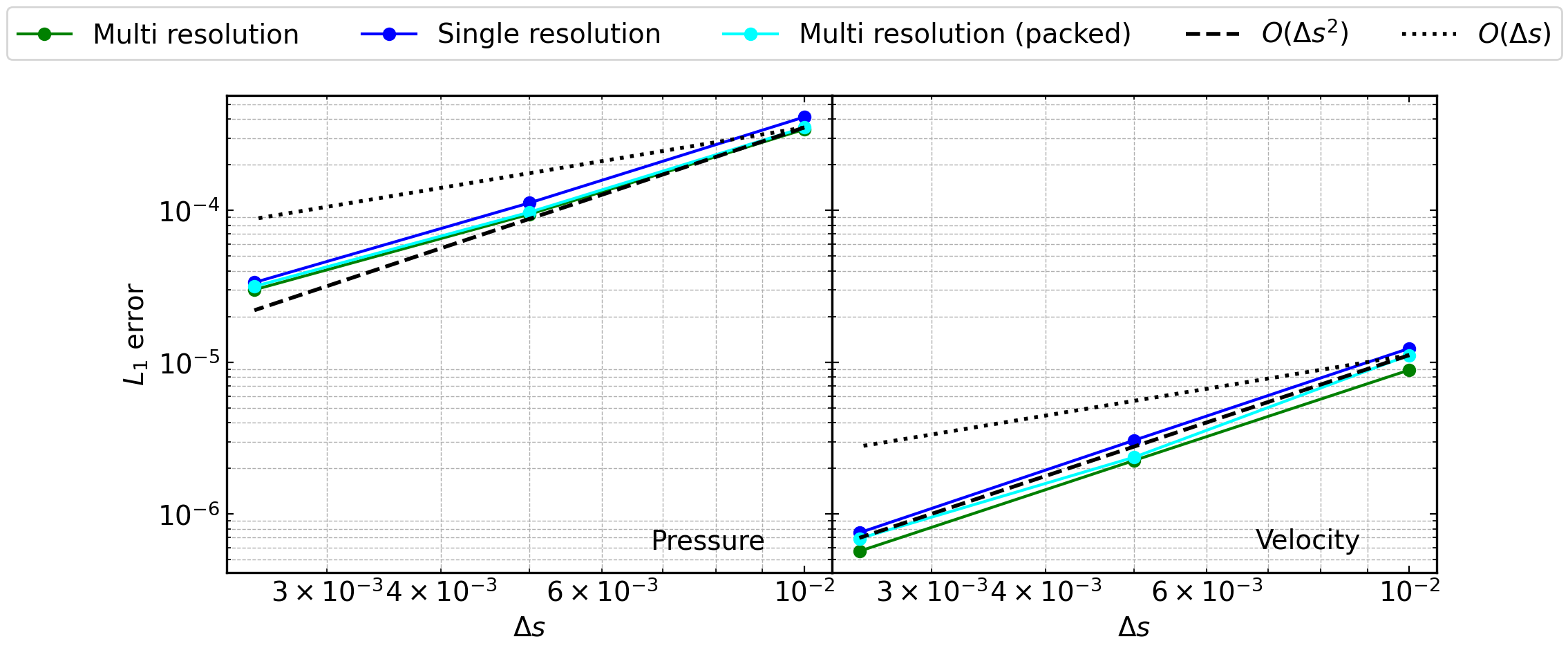}
\caption{$L_1$ error in pressure and velocity after 100 time steps using the
MS in \cref{eq:mms_lap} with $\alpha = 0.5$ and $\nu=0.25$.}
  \label{fig:mms_lap_pst}
\end{figure}

In \cref{fig:mms_lap_pst}, we plot the $L_1$ error for pressure and velocity
after 100 time steps. The proposed scheme demonstrates second-order
convergence in the presence of variable resolution and different particle
distributions.

\subsection{Second-order splitting and merging}
\label{sec:adapt}

In this section, we modify the original splitting and merging technique
proposed by \citet{mutaAdaptive2022} to ensure a second-order convergent
property update for the newly created particles. This enables the present
scheme to remain SOC along with the particle splitting and merging.  However,
the proposed idea is applicable to any adaptive particle refinement method
\cite{vacondio2013variable,yang_adaptive_2019}. For brevity, we refer to
\cite{mutaAdaptive2022} as MAPR (abbreviated for Muta et al. (2022) Adaptive
Particle Refinement) in this discussion. We discuss the proposed modification
here and the complete algorithm is elaborated in
\ref{sec:adaptive-soc_org}. In the next sub-section, we first discuss the
particle splitting technique.

\subsubsection{Splitting a particle}

In MAPR, one particle of mass $m$ is split into seven child particles. The
mass of all the particles is set equal to $m/7$. The properties like velocity
and pressure of the child particles are set equal to the parent
particle. However, in this work we propose to perform Taylor series correction
to determine the property value. Assuming that the particle represents a
scalar field $\phi$, a new particle $\ten{x}_{sp}$ away from the parent
particle uses the first order Taylor-series correction such that
\begin{equation}
  \phi(\ten{x}_s) = \phi(\ten{x}_p) + \ten{x}_{sp} \cdot \left< \nabla \phi(\ten{x}_p) \right>,
  \label{eq:tg_corr_split}
\end{equation}
where $\ten{x}_{sp} = \ten{x}_s - \ten{x}_p$ is the displacement of child
particle $s$ from the parent particle $p$, and $\left< \nabla \phi(\ten{x}_p)
\right>$ is the gradient of the property computed on the parent particle. This
gradient should be computed before the splitting/merging is performed. This is
essential because the original particle distribution represents the original
field variable and can be used to evaluate second-order accurate gradients.

\subsubsection{Iterative Shifting while splitting/merging}

In MAPR, the particles are iteratively shifted in order to make the
particles uniform while splitting and primarily while merging (see algorithm in
\ref{sec:adaptive-soc_org}). We use the shifting technique proposed in
\cite{diff_smoothing_sph:lind:jcp:2009}. The shift vector
\begin{equation}
  \delta \ten{x}_i = \frac{0.5 h_i^2}{\beta_i} \sum_j \left( 1 +0.2
  \left(\frac{W_j}{W(\Delta s)}\right)^4\right) \frac{m_j}{\psi_o} \nabla
  W_j,
  \label{eq:shifting}
\end{equation}
where $W(\Delta s)$ is the smoothing kernel value with the average particle
spacing. We correct the shift vector using the $\beta_i$ term to accommodate
the effect of the change of smoothing length. We note that a kernel gradient
correction should not be employed. We iteratively shift the particles to a
prescribed number of times (thrice in the current work) and correct the
properties using the Taylor series correction such that any property
\begin{equation}
  \phi(\ten{x}_i + \delta \ten{x}_i) = \phi(\ten{x}_i) + \delta \ten{x}_{i} \cdot \left< \nabla
  \phi(\ten{x}_i) \right>,
  \label{eq:tg_shift_corr}
\end{equation}
where  $\left< \nabla \phi(\ten{x}_i) \right>$ is the gradient computed
before the start of the splitting/merging process.

\subsubsection{Merging particles}

In MAPR, merging is performed multiple times in a single split/merge cycle
to enable parallel processing. We note that at a time only two particles merge
to become one and a next set of particles are marked for merging afterwards.
In MAPR, the position of the merged particle $\ten{x}_m$ is set as
\begin{equation}
  \ten{x}_m = \frac{m_a \ten{x}_a + m_b \ten{x}_b}{m_m},
  \label{eq:merge_pos}
\end{equation}
where $m_m = m_a + m_b$ is the mass of the merged particle. The velocity
or pressure of the particle is set as
\begin{equation}
  \phi_m = \frac{m_a \phi_a + m_b \phi_b}{m_m}.
\end{equation}
However, we note that the original particles represent a scalar field
$\phi$ and the resulting merged particle must represent the same field.
Therefore, the property value using a first-order Taylor series
approximation from the nearest particle $n$ to the new merged location,
\begin{equation}
  \phi(\ten{x}_m) = \phi(\ten{x}_{n}) + \ten{x}_{mn} \cdot \left< \nabla \phi(\ten{x}_{n}) \right>,
  \label{eq:merged-phi}
\end{equation}
where $\left< \nabla \phi(\ten{x}_{n}) \right>$ is the gradient evaluated
before the merging cycle begins. 

One adaptive refinement cycle may consist of multiple merges, we demonstrate
the approximation for the scenarios where the merge is performed between two
newly split particles $s_1$ and $s_2$. The property at those particles are
\begin{equation}
  \phi(\ten{x}_{s_1}) = \phi(\ten{x}_{p_1}) +  \ten{x}_{s_1 p_1} \cdot \left< \nabla \phi(\ten{x}_{p_1}) \right>,
  \label{eq:s1}
\end{equation}
and
\begin{equation}
  \phi(\ten{x}_{s_2}) = \phi(\ten{x}_{p_2}) +  \ten{x}_{s_2 p_2} \cdot \left< \nabla \phi(\ten{x}_{p_2}) \right>,
\end{equation}
where $p_1$ and $p_2$ are the parent particle. Let us assume $p_1$ is
closer to the merged location, thus the property for the new merged
particle is given by
\begin{equation}
  \phi(\ten{x}_{m}) = \phi(\ten{x}_{s_1}) +  \ten{x}_{m s_1} \cdot \left< \nabla \phi(\ten{x}_{p_1}) \right>,
\end{equation}
From \cref{eq:s1}, we get
\begin{equation}
  \begin{split}
  \phi(\ten{x}_{m}) & = \phi(\ten{x}_{p_1}) + \ten{x}_{s_1 p_1} \cdot
  \left< \nabla \phi(\ten{x}_{p_1}) \right> +  \ten{x}_{m s_1} \cdot \left<
  \nabla \phi(\ten{x}_{p_1}) \right>\\
  ~ & = \phi(\ten{x}_{p_1}) + \ten{x}_{m p_1} \cdot \left< \nabla
  \phi(\ten{x}_{p_1}) \right> \\
  \end{split}
\end{equation}
In a similar manner, we can show that the particle created from a merged
particle maintains a first-order Taylor-series approximation centered at
chosen nearest parent particle. Hence, the proposed merging is second-order
accurate. We note that we copy the gradient values of the parent particle to
the child particles. In the next section, we apply the proposed scheme to
simulate real flows.

\FloatBarrier%
\section{Results and discussion}
\label{sec:results}

In this section, we simulate two test cases viz.\ the Taylor-Green vortex and
flow past a circular cylinder. We compare the results obtained with the
adaptive SPH scheme proposed by \citet{mutaAdaptive2022} and other established
methods in the literature.

\subsection{Taylor-Green Vortex}

The Taylor-Green vortex is an incompressible flow with an exact solution given by
\begin{equation}
  \begin{split}
    u(x, y, t) &= -U \exp(b t) \cos(2 \pi x) \sin(2 \pi y),\\
    v(x, y, t) &= U \exp(b t) \sin(2 \pi x) \cos(2 \pi y),\\
    p(x, y, t) &= -0.25\,U^2 \exp(2 b t) \cos(2 \pi x) \cos(2 \pi y),
  \end{split}
  \label{eq:tg}
\end{equation}
where $b=-8 \pi^2 / Re$, and $Re=U L / \nu$ is the Reynolds number of the
flow. For the simulation, we consider a square domain of $L=1$ m and maximum
velocity $U=1.0$ ms\textsuperscript{-1}. We set the viscosity $\nu=UL/Re$ and
the artificial speed of sound $c_o=20$ ms\textsuperscript{-1}. We simulate the
problem at the Reynolds numbers $100$ and $1000$. We consider a patch of fluid
in the middle of the domain which has exactly half of the particle spacing of
the larger domain. We simulate the flow for $t=2$ s.

\begin{figure}[htbp]
  \centering
\begin{subfigure}[t]{0.48\linewidth}
  \includegraphics[width=\linewidth]{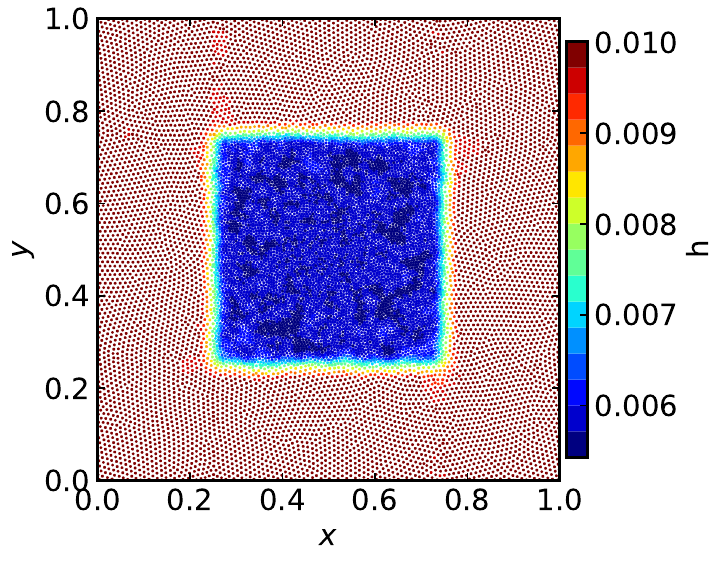}
  \caption{Smoothing length}
\end{subfigure}
\begin{subfigure}[t]{0.48\linewidth}
  \includegraphics[width=\linewidth]{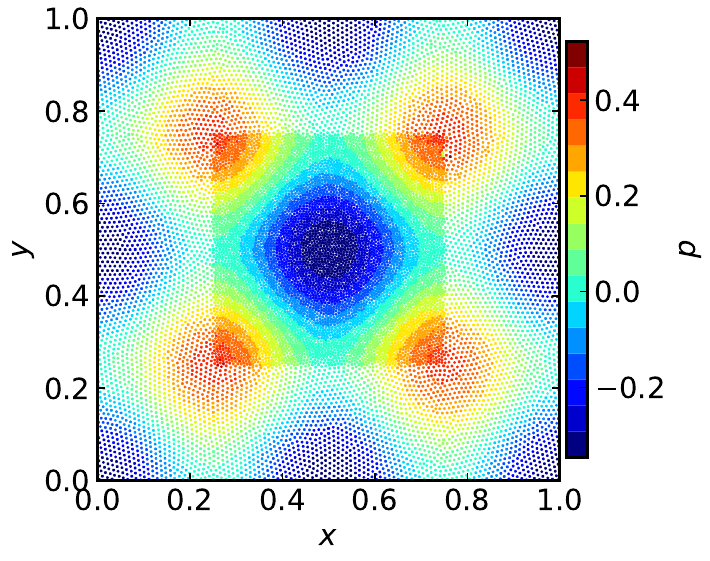}
  \caption{Pressure}
\end{subfigure}
  \caption{Smoothing length variation and pressure at $t=2.0$ s for $Re=1000$.}
  \label{fig:tg_re_1000_p_h}
\end{figure}

In \cref{fig:tg_re_1000_p_h}, we plot the smoothing length variation and
pressure in the domain at $t=2$ s for the $Re=1000$ case. The pressure plot
does not show any error originating at the interface of different resolutions.

\begin{figure}[htbp]
  \centering
  \includegraphics[width=0.5\linewidth]{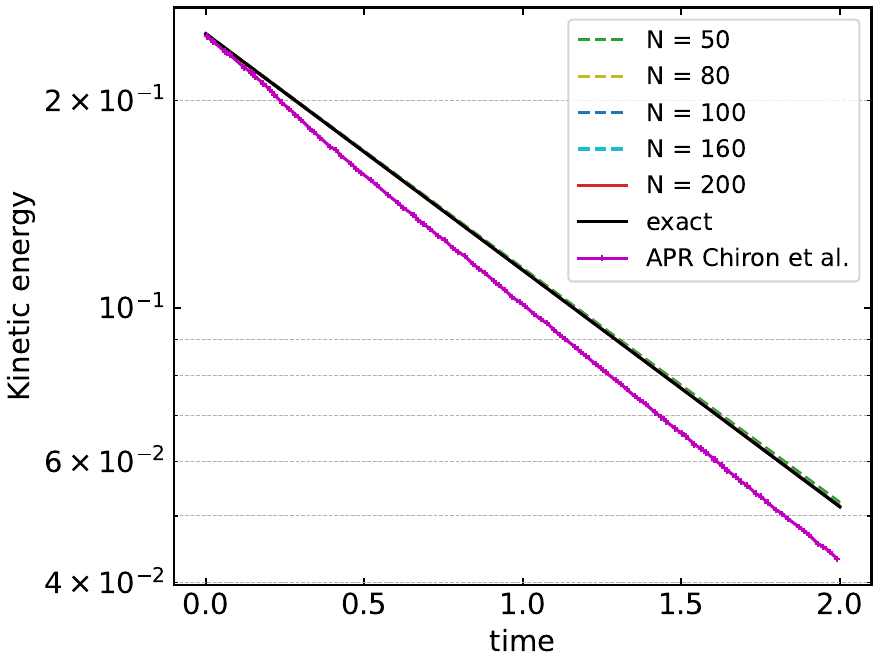}
  \caption{Kinetic energy dissipation of the Taylor-Green vortex for different
resolutions compared with the exact solution and the results of
\citet{chiron_apr_2018}. $N$ represents the number of particles along
$x$-direction indicating the resolution.}
  \label{fig:kin_disp}
\end{figure}

In \cref{fig:kin_disp}, we show the kinetic energy dissipation for the
Taylor-Green vortex and compare the results with that of
\citet{chiron_apr_2018}. The results of the present scheme are shown with 5
different resolutions, viz.~\ $N =50$, $80$, $100$, $160$, and $200$, where
the parameter $N$ is the number particles along the $x$-direction. The
obtained results are very close to the expected exact dissipation rate and are
significantly better than those of \citet{chiron_apr_2018}.

\begin{figure}[htbp]
  \centering
\begin{subfigure}[t]{0.48\linewidth}
  \includegraphics[width=\linewidth]{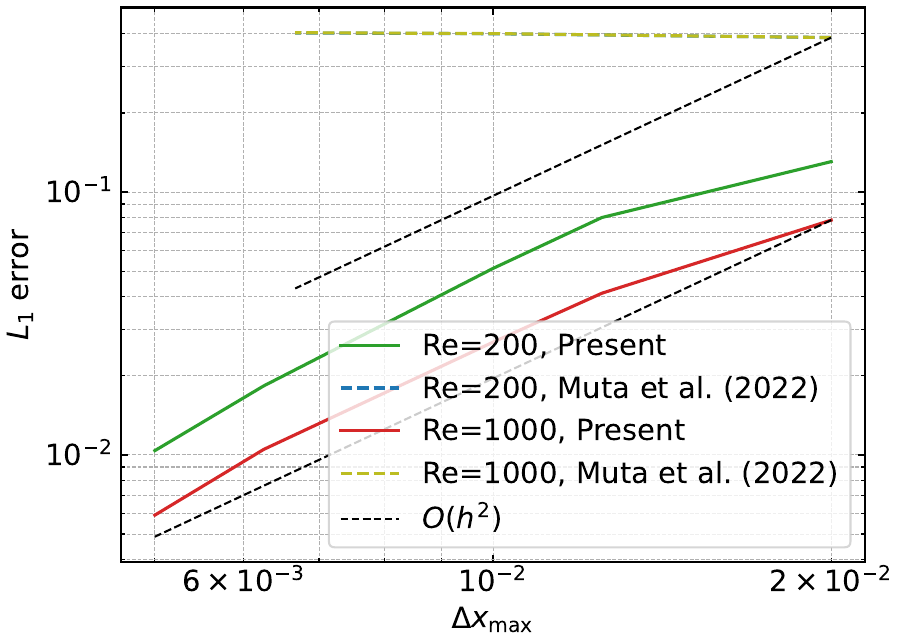}
  \caption{Pressure}
\end{subfigure}
\begin{subfigure}[t]{0.48\linewidth}
  \includegraphics[width=\linewidth]{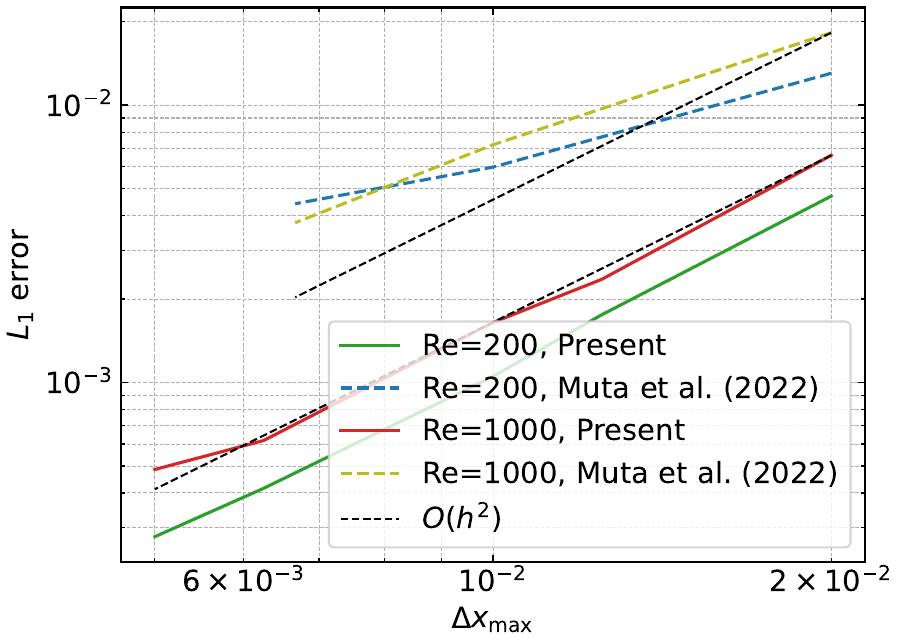}
  \caption{Velocity}
\end{subfigure}
\caption{Comparison of the $L_1$ error in pressure for the adaptive scheme
of \cite{mutaAdaptive2022} and the present scheme.}
\label{fig:tg_conv}
\end{figure}

In \cref{fig:tg_conv}, we plot the convergence in velocity and pressure for
the present method and compare it with the convergence of the results in
\citet{mutaAdaptive2022}. Both velocity and pressure shows second order
convergence. Compared to the results of \cite{mutaAdaptive2022}, the errors
are an order of magnitude lower. Furthermore, convergence in pressure is
entirely absent in their scheme.

\subsection{Flow past a circular cylinder}

The flow past a circular cylinder is a widely used benchmark problem. It is
simple but at the same time exhibits complex flow physics. In this test case,
we consider a two-dimensional circular cylinder of diameter $D=2$ m. The
free-stream velocity $U$ is set to $1$ ms\textsuperscript{-1}. However, the
desired Reynolds number is attained by setting the kinematic viscosity $\nu =
U D / Re$. We consider a uniform particle resolution across the boundaries
i.e. wall, inlet, outlet and the cylinder. Therefore, the boundary condition
implementations suggested in \cite{negiHowTrainYour2022} can be readily
applied to achieve second-order convergence. We apply the boundary condition
implementation proposed by \cite{marrone-deltasph:cmame:2011} for pressure,
slip and no-slip boundary condition. For the inlet and outlet, we employ the
method in \cite{lastiwkanrbc2009} for the non-reflection of the pressure
waves. However, in the case where the reference properties like mean velocity
and pressure are unknown, the method in \cite{neginrbc2020} can be
employed. We also apply non-reflection condition at the far-field inviscid
wall.

\begin{figure}[!htbp]
  \centering
  \includegraphics[width=\linewidth]{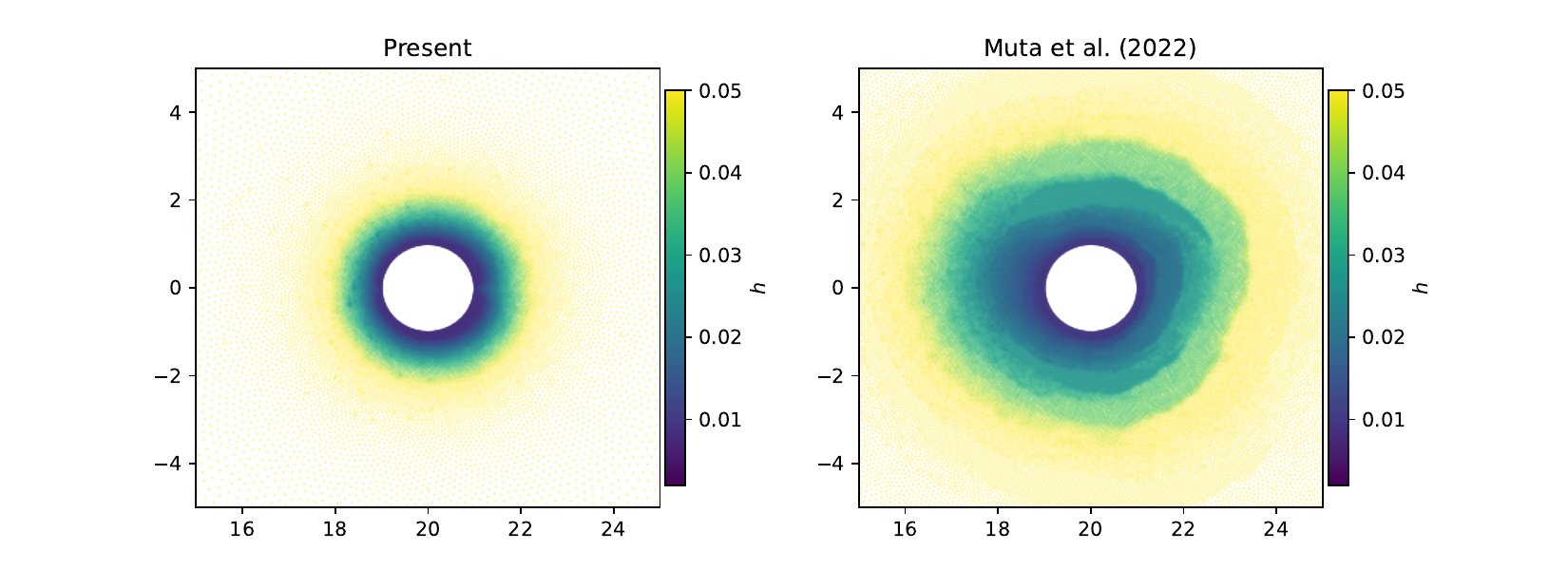}
  \caption{Comparison of the variation of particle resolution near the
cylinder between the present work and \cite{mutaAdaptive2022}.}
  \label{fig:fpc_domain}
\end{figure}

In \cref{fig:fpc_domain}, we show the comparison of particle distribution
around the cylinder for the simulation of $Re=9500$ for the present case and
the one considered in \cite{mutaAdaptive2022}.  We obtain a rapid change in
the particle resolutions due to the improved algorithm in
\cite{haftuParallelAdaptiveWeaklycompressible2022} resulting in a significant
reduction of the total number of particles in the domain. Furthermore, in
\cref{tab:part_comp}, we show the total number of particles at the highest
resolution for different Reynolds numbers.
\begin{table}[!htbp]
  \centering
  \begin{tabular}{c|c|c|c|c}
    \hline
    Re & $C_r$& $D/\Delta s_{min}$ &\multicolumn{2}{c}{Number of particles}\\

    & & &Muta et al. (2022) \cite{mutaAdaptive2022} & Present\\
    \hline
    40 & 1.08 & 500 & 203k & 137k \\
    550 & 1.08 & 500 & 193k & 92k \\
    1000 & 1.08 & 500 & 215k & 89k \\
    3000 & 1.08 & 500 & 212k & 81k \\
    9500 & 1.15 & 1000 & 196k & 141k \\
    \hline
  \end{tabular}
  \caption{Comparison of the total number of particles in the present method
as against that of \citet{mutaAdaptive2022} at the highest resolution at
different Reynolds numbers.}
  \label{tab:part_comp}
\end{table}

For every case, the cylinder is resolved at the highest resolution, $\Delta
s_{\min}$, whereas the inlet/outlet and the wall particles are set at the
lowest resolution $\Delta s_{\max} = 0.5$. In this work we have considered
three different highest resolutions for the cylinder namely, $D/ \Delta
s_{\min} = 100$, 200, 500, while keeping a fixed $\Delta s_{\max}$. The growth
constant $c_r=1.15$. Additionally, we employ solution adaptivity based on
vorticity magnitude to refine the vortices as they develop. For the $Re =
9500$ case, we also employ a highest resolution of $D/ \Delta s_{\min} =
1000$. We demonstrate the qualitative accuracy across the resolution for the
$Re=9500$ case alone.

\begin{figure}[!htpb]
  \centering
  \includegraphics[width=0.9\linewidth]{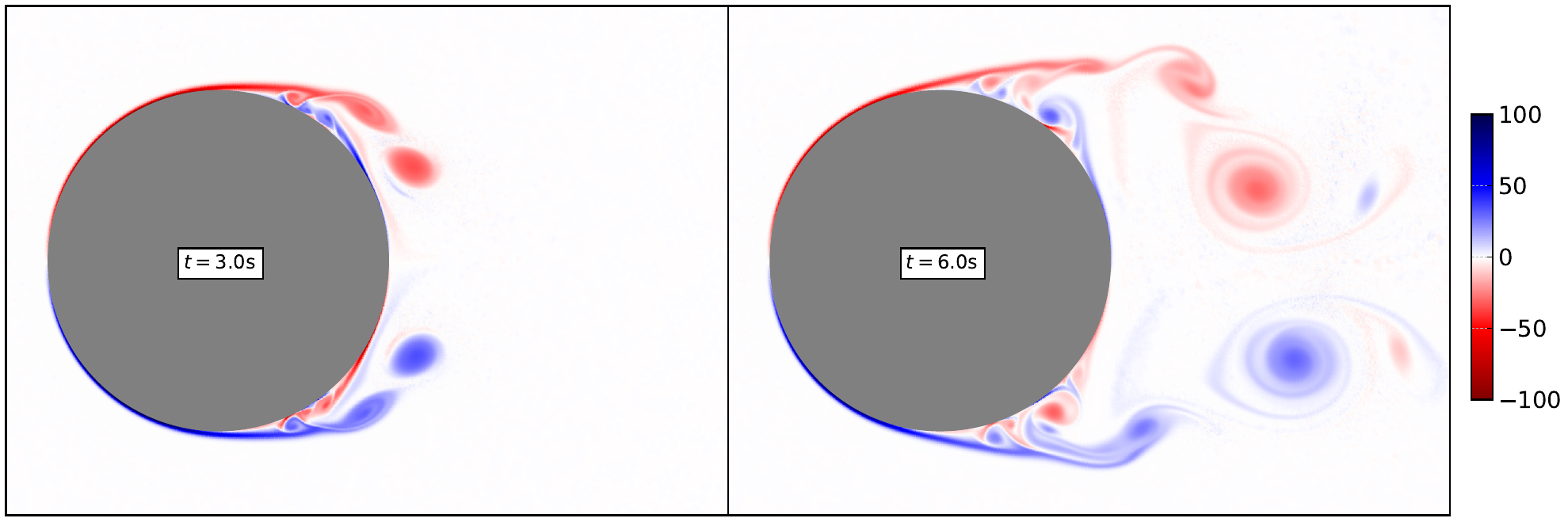}
  \caption{Vorticity distribution for $Re=9500$ at $t = 3$\,s and 6\,s.}
  \label{fig:fpc_vor_9500}
\end{figure}

In \cref{fig:fpc_vor_9500}, we plot the vorticity at $t=3$ s and $6$ s. The
vorticity field shows no jump or variation due to the presence of different
resolution particles in the vicinity of the solid.

\begin{figure}[!htpb]
  \centering
  \includegraphics[width=0.9\linewidth]{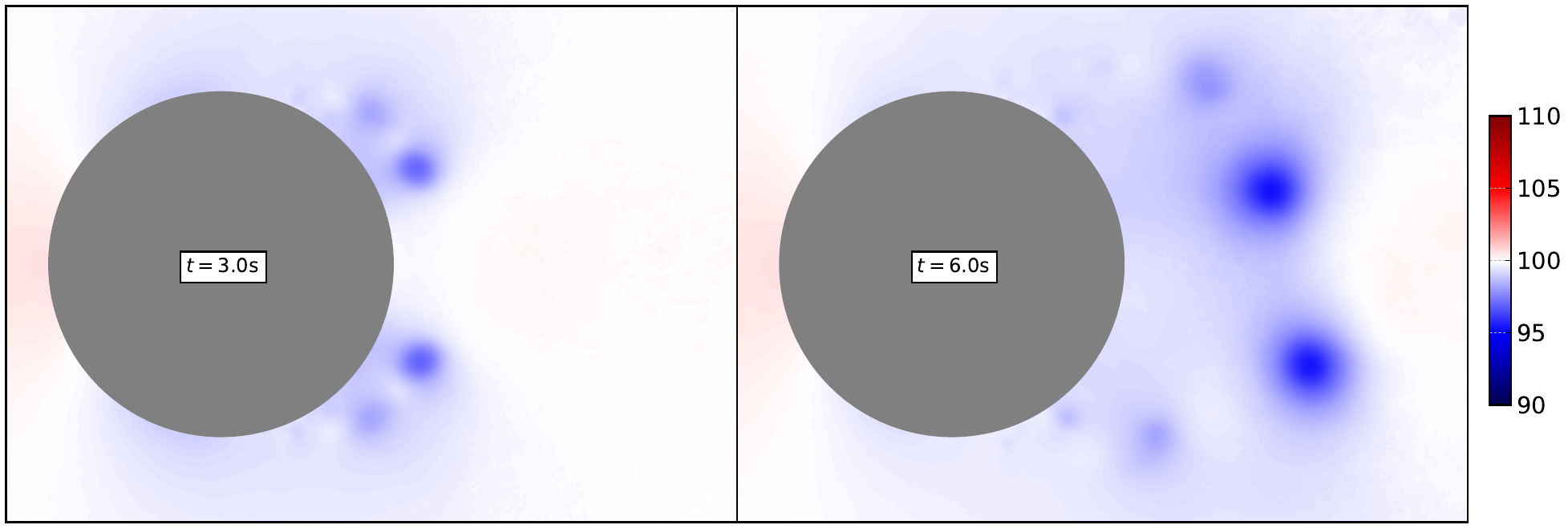}
  \caption{Pressure distribution for $Re=9500$ at $t = 3$\,s and 6\,s.}
  \label{fig:fpc_p_9500}
\end{figure}

In \cref{fig:fpc_p_9500}, we plot the pressure variation around the cylinder
at $t=3$\,s and $6$\,s. Similar to vorticity distribution, the pressure also
does not show any variation due to the change in smoothing
length. Additionally, the pressure remains around $100 Pa$ which is the
reference pressure. It shows that the pressure remains bounded as expected for
a weakly-compressible scheme.

To perform a quantitative comparison, we compute the skin-friction and
pressure drag. To compute the skin-friction drag force $f_{sf}$, we
compute the volume integral of the viscous force in the volume $\Omega$
defined by the boundary of the cylinder, which is defined for a given dynamic
viscosity $\mu$ as
\begin{equation}
  \ten{f}_{sf} = \int_\Omega \mu \nabla^2 \ten{u} d \ten{x}.
\end{equation}
Using the divergence theorem, we can write
\begin{equation}
  \ten{f}_{sf} = \int_{\partial \Omega} \mu \nabla \ten{u} \cdot \ten{n} d
  S(\ten{x}) \approx \sum_{i=1}^N \mu \left<\nabla \ten{u}\right>_i \cdot
  \ten{n}_i S(\ten{x}_i),
\end{equation}
where $S(\ten{x})$ is surface area function, $\ten{n}$ is the outward normal,
$\left<\nabla \ten{u}\right>_i$ is the SPH approximation of the velocity
gradient computed during the simulation. The boundary of the cylinder is
discretized into $N$ equal elements. The skin-friction drag coefficient is
evaluated using the $x$-component of the computed force divided by
$\frac{1}{2} \rho_o U^2$.

\begin{figure}[!htpb]
  \centering
  \includegraphics[width=0.85\linewidth]{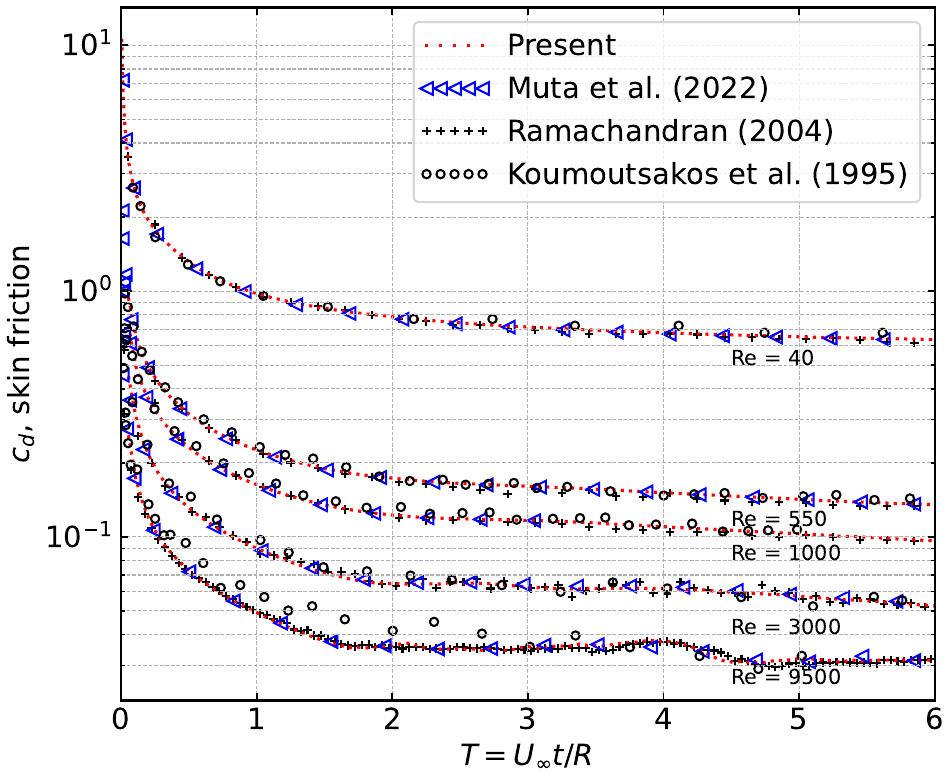}
  \caption{Skin-friction drag history for different Reynolds numbers compared
    with \cite{ramachandran2004,koumoutsakos1995,mutaAdaptive2022}.}
  \label{fig:fpc_drag}
\end{figure}

In \cref{fig:fpc_drag}, we plot the skin-friction drag for the cylinder. The
computed skin-friction drag closely follows the results in
\cite{ramachandran2004,koumoutsakos1995} as well as the adaptive SPH method of
\citet{mutaAdaptive2022}.

In a similar manner, pressure drag force $\ten{f}_{pf}$ is computed using the
force due to pressure gradient in the volume defined by the boundary of the
cylinder, which is given by
\begin{equation}
  \ten{f}_{pf} = \int_\Omega \nabla \cdot (p \mathbf{I}) d \ten{x},
\end{equation}
where $\mathbf{I}$ is an identity matrix. Using the divergence theorem, we
can write
\begin{equation}
  \ten{f}_{sf} = \int_{\partial \Omega} p \mathbf{I} \cdot \ten{n} d
  S(\ten{x}) \approx \sum_i^N \left<p\right>_i \mathbf{I} \cdot
  \ten{n}_i S(\ten{x}_i),
\end{equation}
where $\left<p\right>_i$ is the pressure evaluated using SPH approximation.
The pressure drag coefficient is evaluated by dividing the $x$-component with
$\frac{1}{2} \rho_o U^2$.

\begin{figure}[!htbp]
  \centering
\begin{subfigure}[t]{0.48\linewidth}
  \includegraphics[width=\linewidth]{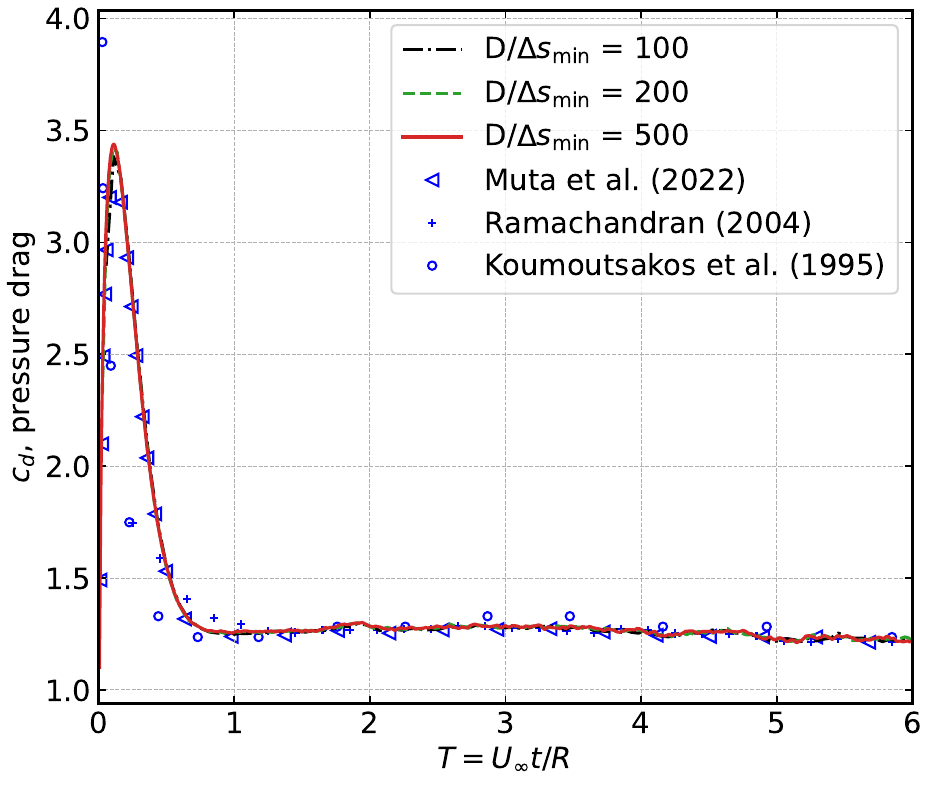}
  \caption{$Re=40$}
\end{subfigure}
\begin{subfigure}[t]{0.48\linewidth}
  \includegraphics[width=\linewidth]{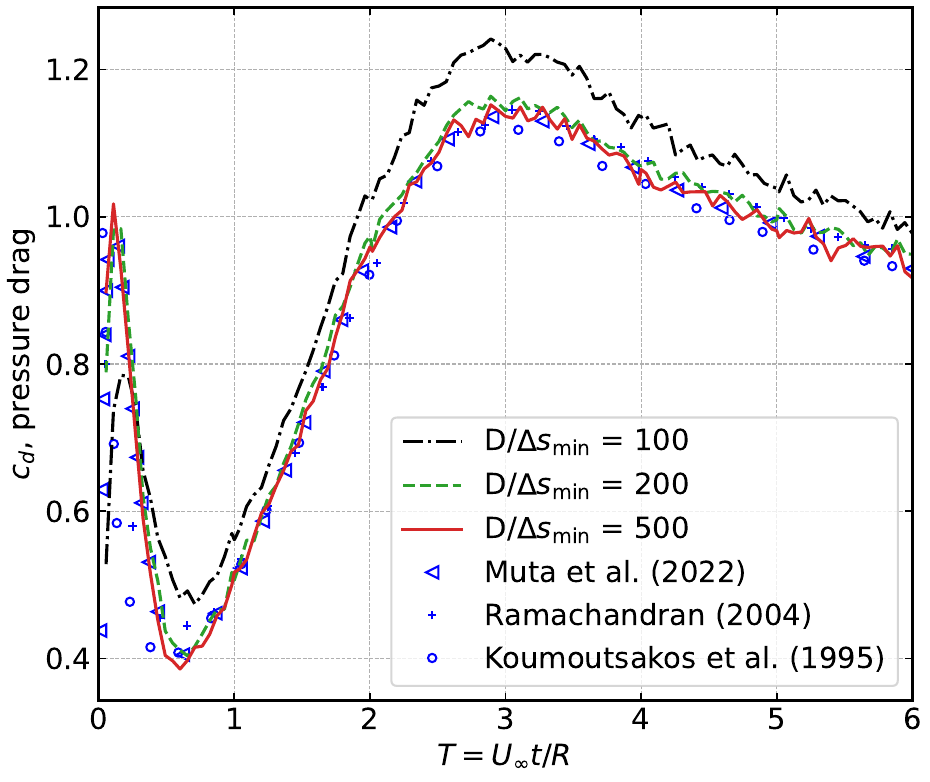}
  \caption{$Re=550$}
\end{subfigure}

\begin{subfigure}[t]{0.48\linewidth}
  \includegraphics[width=\linewidth]{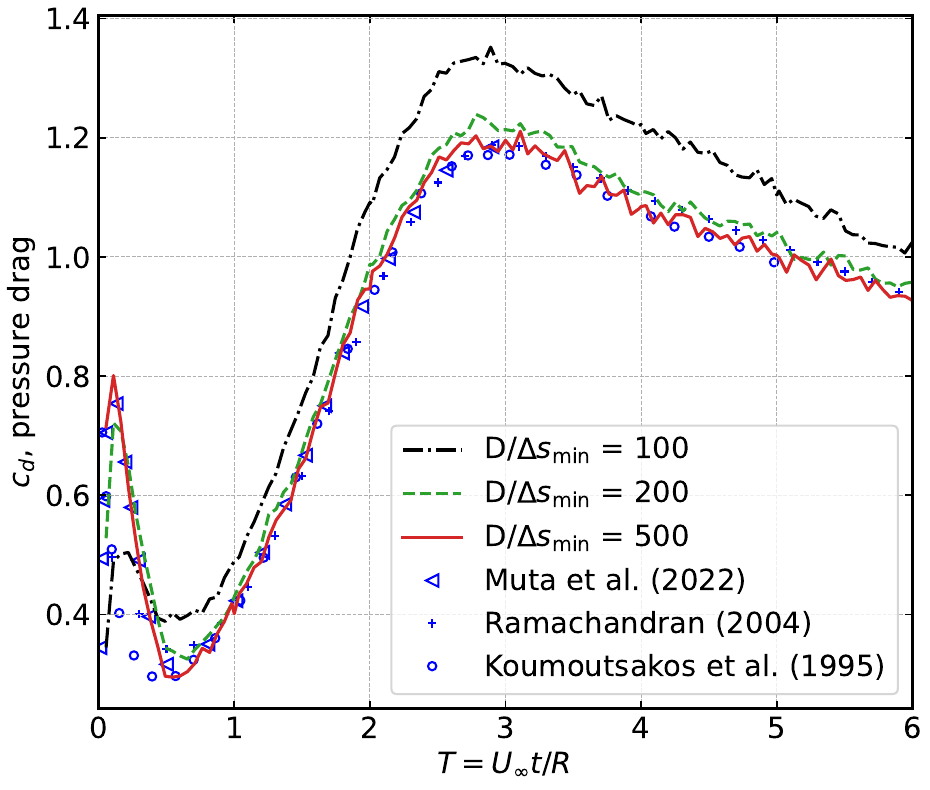}
  \caption{$Re=1000$}
\end{subfigure}
\begin{subfigure}[t]{0.48\linewidth}
  \includegraphics[width=\linewidth]{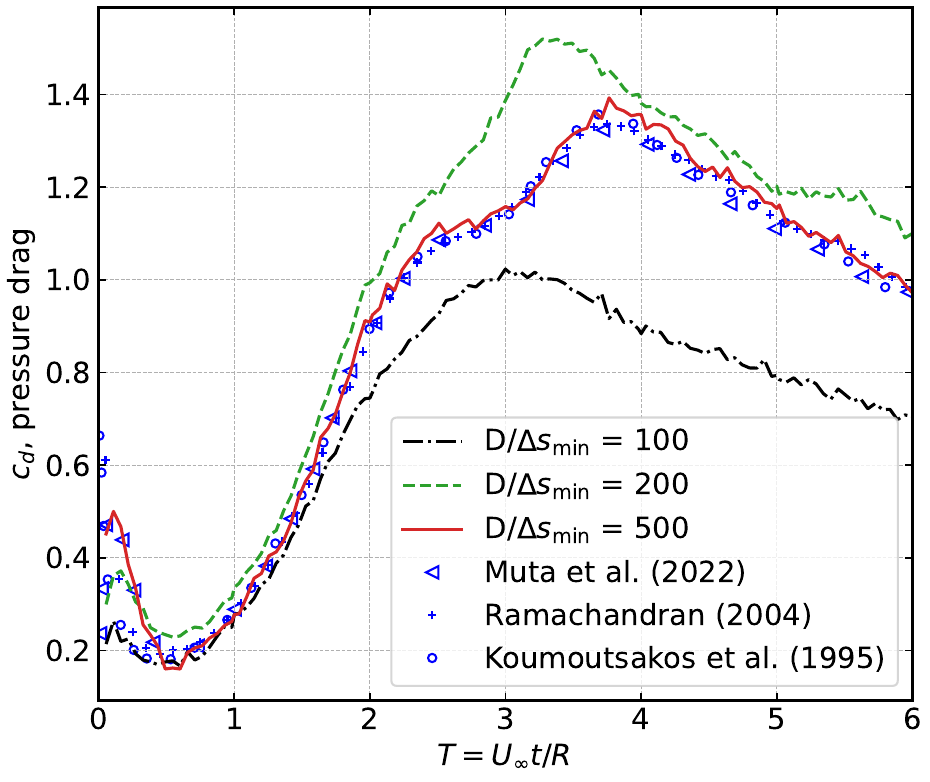}
  \caption{$Re=3000$}
\end{subfigure}
\caption{Coefficient of pressure drag history for different resolutions
compared with the results in
\cite{ramachandran2004,koumoutsakos1995,mutaAdaptive2022}.}
\label{fig:pressure_drag_40_3000}
\end{figure}

In \cref{fig:pressure_drag_40_3000}, we plot the pressure drag
coefficient for $Re=40$, $550$, $1000$, and $3000$ across different
resolutions. We note that the higher Reynolds number flows behave poorly at
lower resolution. However, at resolution $D/\Delta s_{min}=500$ all the cases
match those presented in the literature.

\begin{figure}[!htbp]
  \centering
  \includegraphics[width=0.85\linewidth]{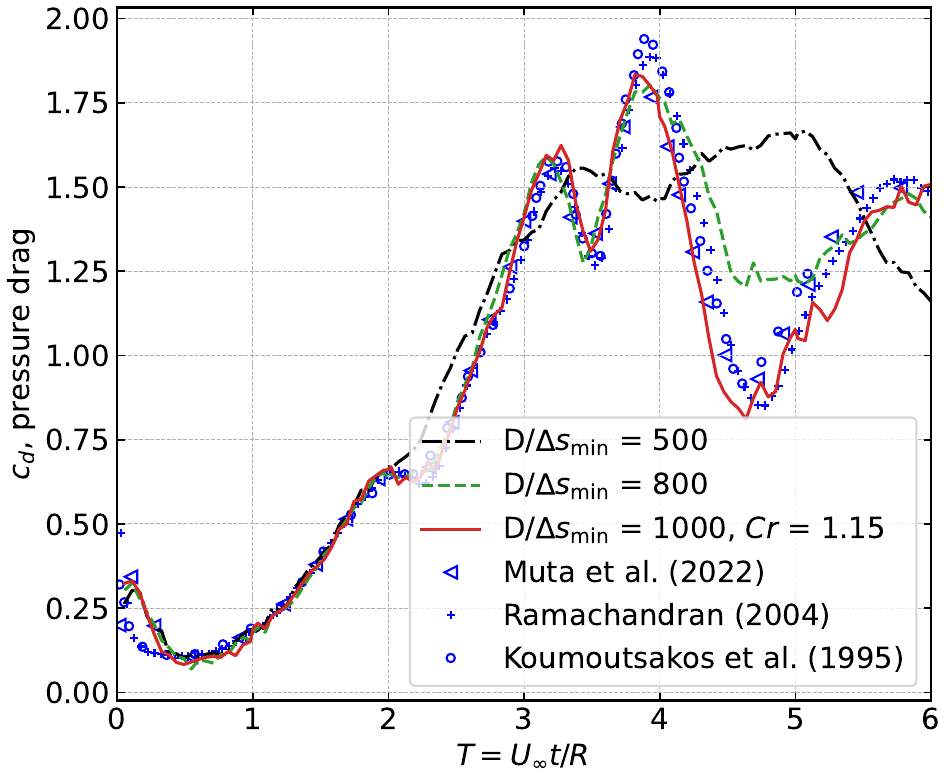}
\caption{Coefficient of pressure drag history for $Re = 9500$ for different
resolutions compared with the results in
\cite{ramachandran2004,koumoutsakos1995,mutaAdaptive2022}.}
\label{fig:drag_9500}
\end{figure}

In \cref{fig:drag_9500}, the $Re=9500$ case, a further increment in the
highest resolution to $D/ \Delta s_{\min} = 1000$ is required to properly
capture the pressure drag. We also use a $C_r$ value of 1.15 to increase the
rate of change of resolution hence, decreasing the total number of
particles. This also demonstrates that the proposed method works with much
rapid change in resolution.

\FloatBarrier%
\section{Conclusions}
\label{sec:conclusions}

In this study, we propose a second-order convergent, weakly-compressible
scheme for adaptive resolution SPH. We propose to employ a second-order kernel
gradient correction in the presence of variable resolution to accurately
evaluate the gradient, divergence and Laplacian of a scalar field. More
importantly, the SPH operator discretizations and the kernel correction
matrices must consistently use one-sided kernel, $W_i = W(\ten{x}_{ij}, h_i)$,
to ensure first-order consistency. We also demonstrate successful application
of the method of manufactured solution to verify the second-order convergence
of the proposed scheme during the development. To maintain second-order
accuracy while adaptively splitting and merging of particles, we employ a
Taylor series correction to consistently update the properties. We emphasise
to compute property gradients before the split/merge process. Using the
Taylor-Green vortex problem, we demonstrate the convergence and accuracy of
the proposed scheme. We simulate the flow past a circular cylinder to
demonstrate the accuracy of skin friction and pressure-drag while using a
significantly lower total number of fluid particles. In the future, we plan to
extend the method to free-surface and multiphase flow problems.

\appendix

\section{Adaptive splitting and merging algorithm}
\label{sec:adaptive-soc_org}

In this section, we highlight the differences and describe the particle
splitting and merging algorithm proposed by
\cite{mutaAdaptive2022,haftuParallelAdaptiveWeaklycompressible2022} for the
sake of completeness. As mentioned in \cref{sec:adapt}, the splitting and
merging processes are not changed in the present method.

The initialization of the fluid particles before the start of the simulation
is described in \cite{haftuParallelAdaptiveWeaklycompressible2022}. However,
any other method to generate a adaptive particle distribution is supported and
does not effect the described SOC algorithm.

In \cref{alg:update-h}, we determine the resolution level $h_{\text{tmp}}$ for
a particle.  For each particle, we compute three quantities viz.\ $\Delta
s_{\text{max}}$, $\Delta s_{\text{min}}$, and geometric average spacing
$\Delta s_{\text{avg}}$ in its neighborhood. For a particle, if the ratio
${\frac{\Delta s_{\max}}{\Delta s_{\min}}} < {(C_r)}^3$ (where, $C_r$ is the
prescribed rate of change of resolution), then it is resolved further to
$\min(\Delta s_{\max}, C_r \Delta s_{\min})$ otherwise the new resolution is
set as the current $\Delta s_{\text{avg}}$. The mass of the particle which is
not fixed is reset as $\psi_i h^{2}_{\text{tmp}}$ with $m_{i, \max}= 1.05
m_i$, and $m_{i, \min} = 0.5 m_i$ required to determine splitting and merging
of particles.
\begin{algorithm}[!htp]
  \caption{Update spacing of particles.}
  \label{alg:update-h}
\begin{algorithmic}[1]
  \State{Define $\Delta s$ for all particles}
 \For{all $i$ of particles which are not fixed}
  \State{find neighbors of $i$}
  \State{find $\Delta s_{\min}$, and $\Delta s_{\max}$ in the neighborhood}
  \State{find average spacing $\Delta s_{\text{avg}}$ in the neighborhood}
  \If{${\frac{\Delta s_{\max}}{\Delta s_{\min}}} < {(C_r)}^3$}
  \State{$h_{\text{tmp}} \leftarrow \min(\Delta s_{\max}, C_r \Delta s_{\min})$}
  \Else{}
  \State{$h_{\text{tmp}} \leftarrow \Delta s_{\text{avg}}$}
  \EndIf{}
  \EndFor{}
  \For{all $i$ of particles which are not fixed}
  \State{$m_i \leftarrow \psi_i h^{2}_{\text{tmp}}$}
  \State{$m_{i, \max} \leftarrow 1.05 m_i \quad m_{i, \min} \leftarrow 0.5 m_i$}
  \EndFor{}
\end{algorithmic}
\end{algorithm}

The particles are split and merged according to the $m_{i, \max}$ and $m_{i,
\min}$ values. The particles are split if the average mass of particle $\sum_j
m_j$ is greater the $m_{i, \max}$. The particles are split into $7$ daughter
particles at a distance $0.4 h$ away from the parent particle in a circle,
with one of the particles at the center. The split particles are initially set
to have their smoothing lengths as $h=0.9 h_i$. In \cref{alg:splitting}, we
describe the splitting procedure in detail. The particle properties are
updated in the lines starting with \textsuperscript{\textdagger} as described
in \cref{sec:adapt} to ensure second-order convergence.
\begin{algorithm}[!htpb]
  \caption{Splitting of the particles.}%
  \label{alg:splitting}
\begin{algorithmic}[1]
  \State{$\epsilon \leftarrow 0.4 \quad \alpha \leftarrow 0.9$}
  \For{all $i$ of particles which are not fixed}
  \If{$m_i > m_{i, \max}$}
  \State{split particle $i$ to 7 daughters}
  \State{$\ten{x}_{i, 0} \leftarrow \ten{x}_i$}
  \For{$k = 1;\ k < 7;\ k\texttt{++}$}
  \State{$\ten{x}_{i, k} \leftarrow \ten{x}_i +
    \epsilon h_i \cos\left(\frac{k\pi}{3}\right)$}
  \EndFor{}
  \For{$k = 0;\ k < 7;\ k\texttt{++}$}
  \State{$\psi_{i, k} \leftarrow \psi_i$}
  \State{$m_{i, k} \leftarrow \frac{m_i}{7}, \quad
  h_{i, k} \leftarrow \alpha h_i$}
  \State{{\textsuperscript{\textdagger}}$p_{i, k} \leftarrow p_i +
(\ten{x}_{i, k} - \ten{x}_{i}) \cdot \nabla p_i$}
  \State{{\textsuperscript{\textdagger}}$\ten{u}_{i, k} \leftarrow \ten{u}_i +
(\ten{x}_{i, k} - \ten{x}_{i}) \cdot \nabla \ten{u}_i$}
  \State{{\textsuperscript{\textdagger}}$\nabla p_{i, k} \leftarrow \nabla p_i$}
  \State{{\textsuperscript{\textdagger}}$\nabla \ten{u}_{i, k} \leftarrow \nabla \ten{u}_i$}
  \EndFor{}
  \EndIf{}
  \EndFor{}
\end{algorithmic}
\end{algorithm}

In \cref{alg:merging}, we show the merging procedure in detail. Two particles
are in a neighborhood of each other are merged, if the merged mass $m_{m}$ is
less than $m_{i, \max}$ of the particle. The merged particle is placed at the
mass-averaged location with the masses of the particles added.  The properties
after merging are updated as descibed in \cref{sec:adapt} to ensure
second-order convergence. After merging, the smoothing length is set as
\cite{vacondio2013variable}
\begin{equation}
  \label{eq:merged-h}
  h_{i} = {\left( \frac{m_m W(\ten{0}, 1)}
      {m_i W(\ten{r}_m - \ten{r}_i, h_i) +
        m_j W(\ten{r}_m - \ten{r}_j, h_j)} \right)}^{1/d},
\end{equation}
where $d$ is the number of spatial dimensions.

\begin{algorithm}[!htp]
  \caption{Merging of particles.}%
  \label{alg:merging}
\begin{algorithmic}[1]
  \For{all $i$ which are not fixed}
  \State{find neighbors of $i$}
  \If{$m_i \le m_{i, \max}$}
  \For{$j$ in neighbor indices}
  \State{$m_{\text{merge}} \leftarrow m_i + m_j$}
  \State{$m_{\max, \min} \leftarrow \min(m_{i, \max}, m_{j, \max})$}
  \If{$m_{\text{merge}} < m_{\max, \min}$ \textbf{\&} $j$ is closest
    of all neighbors}
  \State{store index $j$ for merging}
  \EndIf{}
  \EndFor{}
  \EndIf{}
  \EndFor{}
  \For{all $i$ of particles which are to be merged}
  \If{merge pair of $i$ is $j$ and merge pair of $j$ is $i$ and $i < j$}
  \State{update $\ten{x}_i$ using \cref{eq:merge_pos}}
  \State{{\textsuperscript{\textdagger}}update $\ten{u}_i$, $p_i$ using
\cref{eq:merged-phi}}
  \State{{\textsuperscript{\textdagger}}copy gradient properties, $\nabla
\ten{u}_i$ and $\nabla p_i$}
  \State{update $h_i$ with its merged pair index using \cref{eq:merged-h}}
  \Else
  \If{merge pairs match and $i > j$}
  \State{Delete particle $i$}
  \EndIf{}
  \EndIf{}
  \EndFor{}
\end{algorithmic}
\end{algorithm}

\Cref{alg:apr} discusses the overall adaptive procedure in detail. In one
round of splitting and merging operation, the splitting is performed once and
merging is performed thrice to ensure a uniform distribution of mass. After
the splitting and merging, particle shifting is performed as discussed in
\cref{sec:adapt} to ensure second-order convergence.
\begin{algorithm}[!htpb]
  \caption{Adaptive particle refinement procedure.}%
  \label{alg:apr}
\begin{algorithmic}[1]
  \While{$t < t_{\text{final}}$}
  \For{all fluids $i$}
  \State{{\textsuperscript{\textdagger}}Compute gradients, $\nabla \ten{u}$,
    and $\nabla {p}$.}
  \If{fluid particles closest to boundary}
  \State{$\Delta s_i \leftarrow \Delta s_{\min}$}
  \EndIf{}
  \If{fluid particles satisfy solution adaptive criteria}
  \State{$\Delta s_i \leftarrow \Delta s_{\min}$}
  \EndIf{}
  \EndFor{}
  \State{Update the spacing of particles (\cref{alg:update-h})}
  \State{Split the particles (\cref{alg:splitting})}
  \For{$i = 0; i < 3; i\texttt{++}$}
  \State{Merge the particles (\cref{alg:merging})}
  \EndFor{}
  \State{update the smoothing length using \cref{eq:optimal-h}}
  \For{$i = 0; i < 3; i\texttt{++}$}
  \State{shift the particles using \cref{eq:shifting}}
  \EndFor{}
  \State{{\textsuperscript{\textdagger}}correct the particle properties using
\cref{eq:tg_shift_corr}}
  \State{resume the simulation using adaptive SOC scheme (\cref{sec:soc_edac})}
  \EndWhile{}
\end{algorithmic}
\end{algorithm}

Before the start of the SOC scheme and after every splitting and merging
operation, we recompute the particle smoothing length of all the particles
using
\begin{equation}
  \label{eq:optimal-h}
  h_i = C {\left(\frac{1}{\psi_o} \frac{\sum_j m_j}{N_i}\right)}^{1/d},
\end{equation}
where $C = 1.2$ is a constant corresponding to the value of smoothing length factor
used in the simulation.

\FloatBarrier%
\bibliographystyle{model6-num-names}
\bibliography{references}

\end{document}